\documentclass[reprint,twocolumn,showpacs]{revtex4}
\usepackage{amsmath}
\usepackage{amssymb}
\usepackage{color}
\usepackage[pdftex]{graphicx}
\usepackage{hyperref}
\usepackage[caption=false]{subfig}
\usepackage{hyperref}
\begin{document}
\title{Wannier functions using a discrete variable representation for optical lattices}
\author{Saurabh Paul}
\affiliation{Joint Center for Quantum Information and Computer Science, Joint Quantum Institute and University of Maryland, Maryland 20742, USA}
\author{Eite Tiesinga}
\affiliation{Joint Quantum Institute and Joint Center for Quantum Information and Computer Science, National Institute of Standards and Technology and University of Maryland, Gaithersburg, Maryland 20899, USA}
\date{\today}
\begin{abstract}
We propose a numerical method using the discrete variable representation (DVR) for constructing real-valued Wannier functions localized in a unit cell for both symmetric and asymmetric periodic potentials. We apply these results to finding Wannier functions for ultracold atoms trapped in laser-generated optical lattices.
Following Kivelson \cite{kivelson_wannier_1982}, for a symmetric lattice with inversion symmetry, we construct Wannier functions as eigen states of the position operators $\hat x$, $\hat y$ and $\hat z$ restricted to single-particle Bloch functions belonging to one or more bands. To ensure that the Wannier functions are real-valued, we numerically obtain the band structure and real-valued eigen states using a uniform Fourier grid DVR. We then show by a comparison of tunneling energies, that the Wannier functions are accurate for both inversion symmetric and asymmetric potentials to better than ten significant digits when using double-precision arithmetic. The calculations are performed for an optical lattice with double-wells per unit cell with tunable asymmetry along the $x$ axis and a single sinusoidal potential along the perpendicular directions. Localized functions at the two potential minima within each unit cell are similarly constructed, but using a superposition of single-particle solutions from the two lowest bands. We finally use these localized basis functions to determine the two-body interaction energies in the Bose-Hubbard (BH) model, and show the dependence of these energies on lattice asymmetry.
\end{abstract}
\maketitle

\begin{section}{Introduction}

  \begin{figure}
    \centering
    \subfloat[Part 1][]{\includegraphics[width=1.85in]{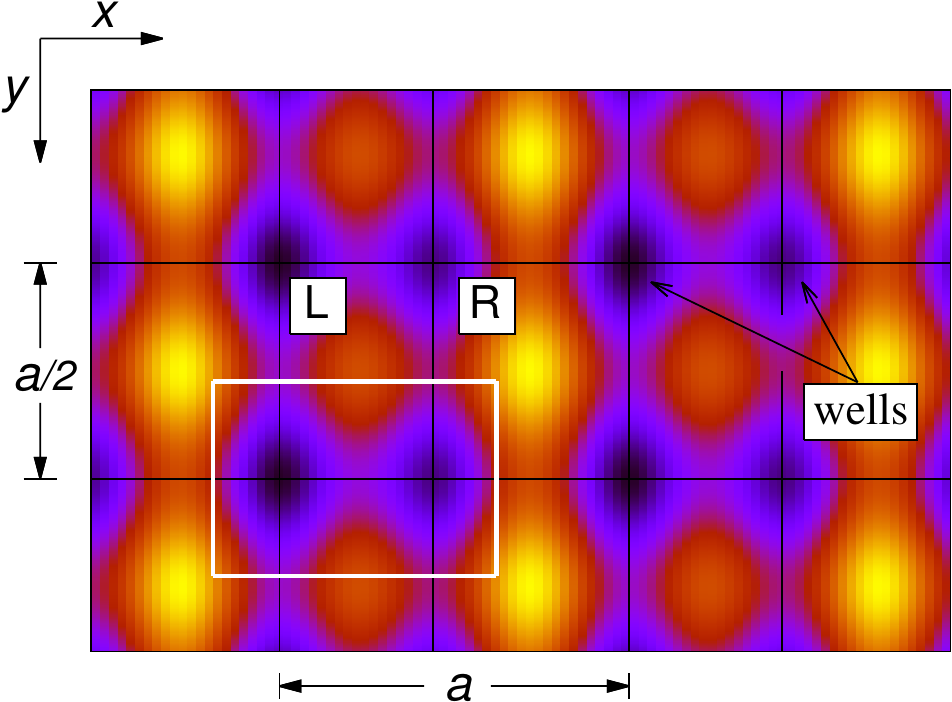}\label{fig:contour}}
    \hspace{2mm}
    \subfloat[Part 2][]{\includegraphics[width=1.35in]{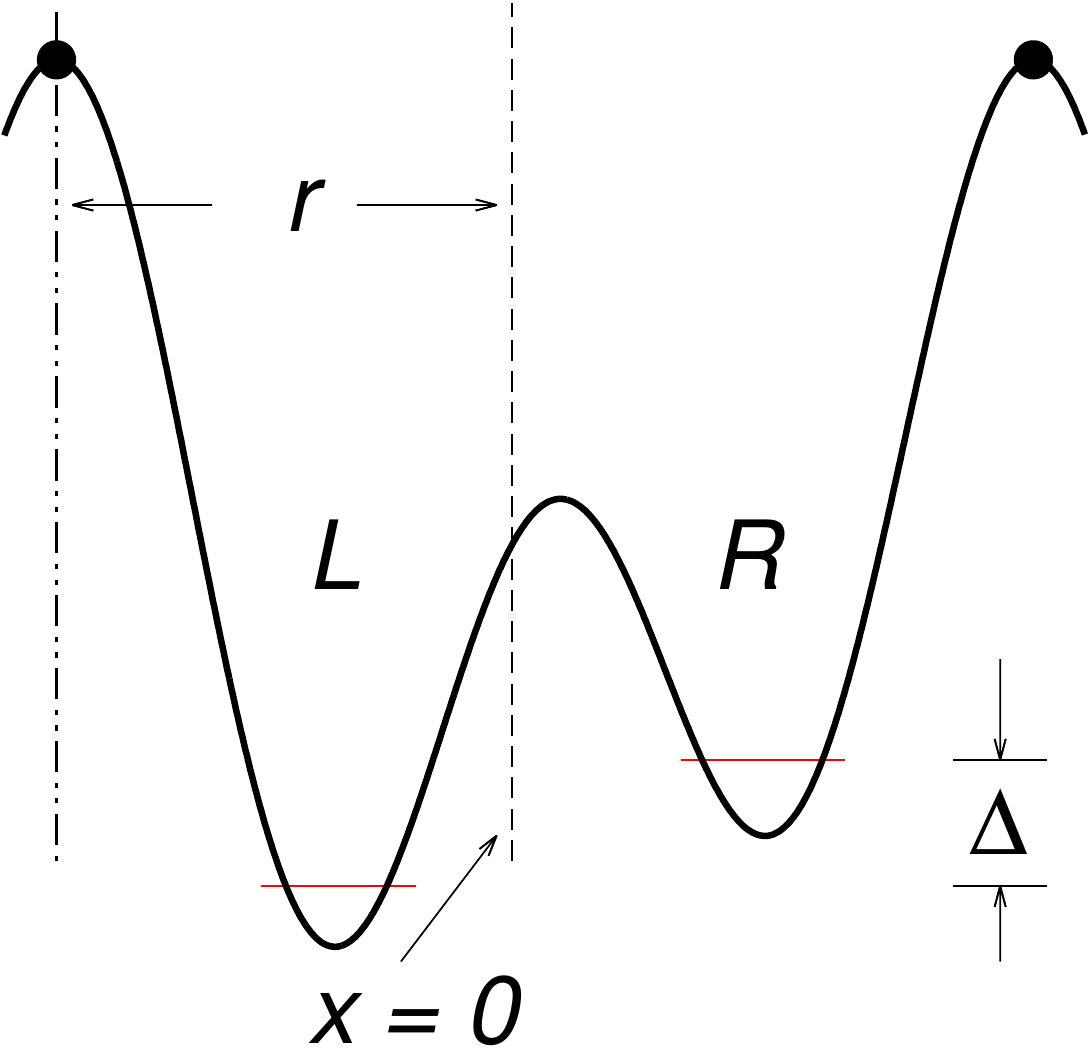}\label{fig:double-well}}
    \caption{(color online) (a) Contour plot of the optical lattice potential in the $xy$ plane, where the potential minima are in dark blue. The white box encloses a unit cell of length $a$ and $a/2$ along $x$ and $y$, respectively. Each unit cell has a double well along the $x$ axis, labeled $L$ and $R$, and a single well along the $y$ and $z$ axes. (b) An asymmetric double-well potential (black curve) as a function of $x$ for $V_1/V_0=1.3$ and $k_Lb=0.21\pi$. The horizontal red curves in the $L$ and $R$ wells represent the lowest two single-particle energy levels. The energy gap between these levels is $\Delta$. The separation between the black dots is the lattice period $a$. The origin $x=0$ of our coordinate system is indicated by the dashed line. For a symmetric lattice, the origin lies on the top of the barrier between the $L$ and $R$ wells. The distance between the origin and the left black dot is $r$.
}
  \end{figure}

  Ultracold atoms in optical lattices form highly tunable systems and are increasingly used to simulate complex quantum many-body Hamiltonians \cite{bloch_many-body_2008,jaksch_cold_2005}. The now very commonly used Bose-Hubbard (BH) model
was first proposed in the context of cold atoms by Ref.~\cite{jaksch_cold_1998}, and its interaction driven quantum phase transition in a cubic lattice was subsequently observed in \cite{greiner_quantum_2002}. Since then, more exotic lattice geometries such as double-well lattices \cite{sebby-strabley_lattice_2006,lee_sublattice_2007,trotzky_time-resolved_2008,atala_direct_2013}, honeycomb, triangular and Kagome lattices \cite{tarruell_creating_2012,jo_ultracold_2012}, and artificial graphene \cite{lee_ultracold_2009,uehlinger_artificial_2013} have been experimentally realized. This has vastly expanded the standard BH model to include additional terms ranging from excited band contributions, beyond nearest-neighbor tunneling to richer on-site and off-site atom-atom interactions \cite{jurgensen_density-induced_2012,luhmann_multi-orbital_2012,bissbort_effective_2012,paul_large_2015}. In conjunction, there has been a growing need to quantitatively model these systems with greater accuracy.

The BH models are an approximation to the full many-body Hamiltonian in the tight-binding (TB) limit, and are written in a single-particle basis of spatially localized wave functions, generally referred to as Wannier functions. The parameters of the BH model are obtained as integrals over these functions. Thus, the key to accurately model these systems is to first construct a set of properly localized orthonormal basis functions. For simple cubic lattices with inversion symmetry, the standard procedure is to construct Wannier functions as ``simple'' superpositions of the Bloch functions belonging to a {\it single} energy band \cite{kohn_analytic_1959,wannier_dynamics_1962}. For more complex lattice geometries with either asymmetries or quasi-degenerate energy bands, this procedure, however, does not lead to basis functions localized at the potential minima within each unit cell. 

A common approach to ensuring localized Wannier functions for atoms in optical lattices is to use non-orthogonal atomic orbitals, modeled as harmonic oscillator wave functions near the potential minima \cite{lee_ultracold_2009,paul_formation_2013}. This underestimates the tunneling energies even for deep lattices where the harmonic approximation is expected to work better. A more general approach developed within the solid-state community is due to Marzari and Vanderbilt \cite{marzari_maximally_1997,marzari_maximally_2012}, where maximally localized  Wannier functions are constructed by minimizing its spread by a suitable gauge transformation of the composite Bloch functions. This scheme has been adapted for atoms in optical lattices \cite{vaucher_fast_2007,modugno_maximally_2012,ibanez-azpiroz_self-consistent_2013,walters_textitab_2013,
ibanez-azpiroz_tight-binding_2013}. Wannier functions obtained using this method, however, are not guaranteed to be real-valued and in turn depend on the choice of gauge transformation. An alternate method for constructing Wannier functions is by minimization of density-induced tunneling and density-density interactions between neighboring unit cells \cite{luhmann_quantum_2014}.

In this paper, we propose an alternative numerical scheme for constructing real-valued Wannier functions. Following Kivelson \cite{kivelson_wannier_1982} who showed that for an inversion symmetric lattice, Wannier functions are eigen states of the position operator, we construct Wannier functions by diagonalization of the position operator expressed in the eigen states of the single-particle Hamiltonian. The localized functions are remarkably accurate even for lattices with a large asymmetry. To ensure that the Wannier functions are strictly real-valued, we obtain the band structure and corresponding real-valued eigen functions using a uniform Fourier-grid discrete variable representation (DVR) \cite{colbert_novel_1992}. General background on the DVR method can be found in \cite{szalay_discrete_1993,tiesinga_photoassociative_1998,littlejohn_general_2002,szalay_one-dimensional_2003}, and some of their uses in ultracold atomic systems can be found in \cite{nygaard_vortex_2004,wall_effective_2015}. Generalized Wannier functions localized at the potential minima in a unit cell are similarly constructed using a superposition of Bloch functions of multiple bands.

The proposed method doesn't suffer from the problems of local minima, as is sometimes the case with the Marzari and Vanderbilt approach of constructing Wannier functions \cite{walters_textitab_2013}. In addition, using the DVR approach intrinsically ensures that the Wannier functions are real-valued. This  differs from the alternative method which uses time reversal symmetry to construct real-valued single-particle basis functions using a superposition of Bloch functions of opposite quasi-momenta \cite{uehlinger_artificial_2013}.

  The remainder of the paper is setup as follows. In Sec.~\ref{sec:OL}, we introduce the asymmetric double-well optical lattice potential, for which we describe the numerical procedure to obtain real-valued DVR-based Wannier functions. The method can be used for general lattices. For clarity, we focus on a particular lattice potential.
In Sec.~\ref{sec:DVR}, we discuss how the single-particle band structure for this lattice can be obtained using a DVR, and also how it compares with that of a plane-wave basis calculation. We also describe how to extend our approach to general lattices. In Sec.~\ref{sec:wfn} real-valued Wannier and localized functions within a double-well potential are obtained using the eigen vectors from the DVR calculations. In Sec.~\ref{sec:validity}, we discuss the accuracy of these numerically obtained Wannier functions by comparing the tunneling energies obtained using these functions to those obtained using a Fourier transform of the band dispersion. We use these DVR-based Wannier functions in Sec.~\ref{sec:int} to compute the two-body interaction energies for various asymmetries. We conclude in Sec.~\ref{sec:conclude}.
\end{section}

\begin{section}{optical lattice potential}
  \label{sec:OL}

  We consider optical lattice potentials that have a double-well structure along the $x$ axis, and a single-well structure along the perpendicular $y$ and $z$ axes. Such a lattice can be constructed using a laser with wave vector $k_L$ and its first harmonic. The potential is given by
  \begin{align}
    \label{eq:potential}
    V(\vec x)&=-V_0\cos^2(k_Lx)-V_1\cos^2\left[2k_L(x+b)\right]\nonumber\\
        &\quad\quad\,-V_2\left[\cos^2(2k_Ly)+\cos^2(2k_Lz)\right],
  \end{align}
where $V_{0,1,2}>0$ are lattice depths. The lattice has periodicity  $a=\pi/k_L$ along the $x$ axis and $a/2$ along the perpendicular directions. The displacement $b$ determines whether the lattice has an inversion-symmetric or asymmetric double-well. It is inversion symmetric for $k_Lb=\pi/4$ and asymmetric otherwise. Throughout, we express energies in units of the recoil energy $E_R=\hbar^2k_L^2/(2m_a)$, where $m_a$ is the atomic mass. Figure~\ref{fig:contour} shows a contour plot of the optical-lattice potential in the $xy$ plane for $k_Lb=0.21\pi$, while Fig.~\ref{fig:double-well} shows the corresponding asymmetric double-well along the $x$ axis. We will concentrate on the potential along $x$ axis in subsequent sections. The perpendicular directions will be needed when estimating two-body interaction energies in Sec.~\ref{sec:int}. 
\end{section}

\begin{section}{Band structure using a discrete variable representation (DVR)}
  \label{sec:DVR}

  \begin{figure}
    \centering
    \includegraphics[width=3.2in]{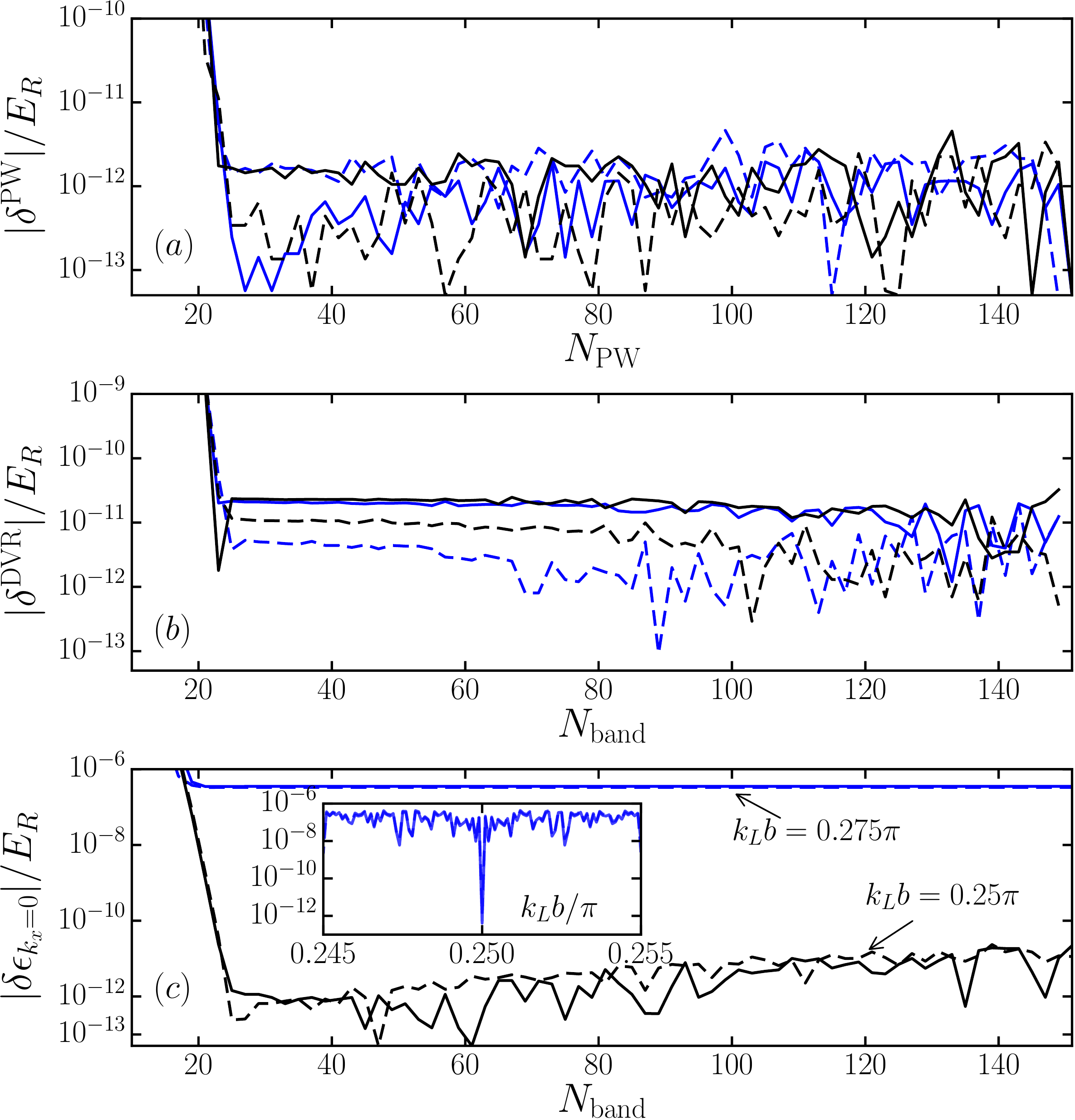}
    \caption{(color online) Panel (a) shows the convergence of the energy dispersion $\epsilon_{\alpha}(k_x)$ at quasi-momentum $k_x=0$ computed with a plane-wave (PW) basis as a function of $N_{\rm PW}$, the number of plane-waves. We plot the difference $\delta^{\rm PW}(N_{\rm PW})=\epsilon_{\alpha}(k_x=0;N_{\rm PW})-\epsilon_{\alpha}(k_x=0;N_{\rm PW}^{\rm max})$, where $N_{\rm PW}^{\rm max}=151$. Panel (b) shows a similar convergence plot using a discrete variable representation (DVR) basis as a function of $N_{\rm band}$, the number of grid points in a unit cell. Plotted is $\delta^{\rm DVR}(N_{\rm band})=\epsilon_{\alpha}(k_x=0;N_{\rm band})-\epsilon_{\alpha}(k_x=0;N_{\rm band}^{\rm max})$, where $N_{\rm band}^{\rm max}=151$ and $M_x=3$. Panel (c) shows a comparison of $\epsilon_{\alpha}(k_x)$ at $k_x=0$ obtained using the DVR and PW basis. We plot $\delta(N_{\rm band})=\epsilon_{\alpha}(k_x=0;{\rm DVR})-\epsilon_{\alpha}(k_x=0;{\rm PW})$ as a function of $N_{\rm band}$. The PW results are obtained with $151$ basis vectors. For all panels black and blue curves are for a symmetric lattice with $k_Lb=\pi/4$ and asymmetric lattice with $k_Lb=0.275\pi$, respectively. Solid and dashed lines correspond to bands $\alpha=1$ and $2$, respectively and lattice depths are $V_0=35E_R$ and $V_1/V_0=1.3$, where $E_R$ is the recoil energy. The inset in panel (c) compares the DVR and PW results as a function of lattice asymmetry $k_Lb$ for fixed $N_{\rm band}=N_{\rm PW}=51$.}
    \label{fig:DVRvsPW}
  \end{figure}

  The single-particle band structure of a periodic potential is generally numerically determined in a plane-wave (PW) basis. For asymmetric lattices, the eigen vectors or the Bloch functions in this basis are complex valued and corresponding Wannier functions are complex as well. We use a discrete variable representation (DVR) to obtain real-valued eigen functions. 
  
  We begin the procedure by discussing the one-dimensional DVR along the $x$ axis. We are interested in solutions that have periodic boundary condition over $M_x$ unit cells. For our double-well potential, it is convenient to apply the shift $x\to x-r$ such that the origin of the $x$ axis coincides with the top of the potential barrier (see Fig.~\ref{fig:double-well}), and consider the interval $( 0,M_xa)$. For a symmetric double-well $r=a/2$, while in general, it  depends on the symmetry parameter $b$. We now introduce the uniformly spaced Fourier grid \cite{colbert_novel_1992}, based on $2N_x+1$ periodic orthonormal basis functions $\phi_n(x) = \exp[i2\pi nx/(M_xa)]/\sqrt{M_xa}$ for $n=0,\pm 1, \ldots, \pm N_x$. Orthonormal DVR  basis functions are $f_i(x)=\langle x|x_i\rangle = \sqrt{\Delta x}\sum_{n=-N_x}^{N_x}\phi_n^*(x_i)\phi_n(x)$, labeled by grid points $x_i = i\Delta x$ with $i=1,\ldots, 2N_x+1$ and $\Delta x=M_xa/(2N_x+1)$. A function $\langle x|x_i\rangle$ is localized around $x_i$ and can be simplified with some trigonometry.

In this representation of grid points, the kinetic energy operator is $T_{ii'}=\langle x_i|T|x_{i'}\rangle$ where
\begin{align}
  \label{eq:dvr2}
  T_{ii'}  & = (-1)^{i'-i}E_R\left(\frac{2\pi}{M_xk_La}\right)^2\nonumber\\
  & \quad\times
    \begin{cases}
      N_x(N_x+1)/3, & i=i',\\
      \dfrac{\cos\left[\pi(i'-i)/(2N_x+1)\right]}{2\sin^2\left[\pi(i'-i)/(2N_x+1)\right]}, & i\ne i',
    \end{cases}
\end{align}
 and to a good approximation the potential energy operator is $\langle x_i|V|x_{i'}\rangle=V(x_i)\delta_{ii'}$ with Kronecker-delta $\delta_{ij}$. In fact, it is this approximation that will limit our numerical accuracy using the DVR. On the other hand, the single-particle Hamiltonian $H_0=T+V$ is a real symmetric matrix for both symmetric and asymmetric lattice potentials and its eigen functions can always be obtained using real arithmetic. We note that in a PW basis, the Hamiltonian for an asymmetric lattice is a complex Hermitian matrix. Typically, we require that $2N_x+1\gg M_x$ leading to many grid points per unit cell.

  \begin{figure*}
    \centering
    \includegraphics[width=6.8in]{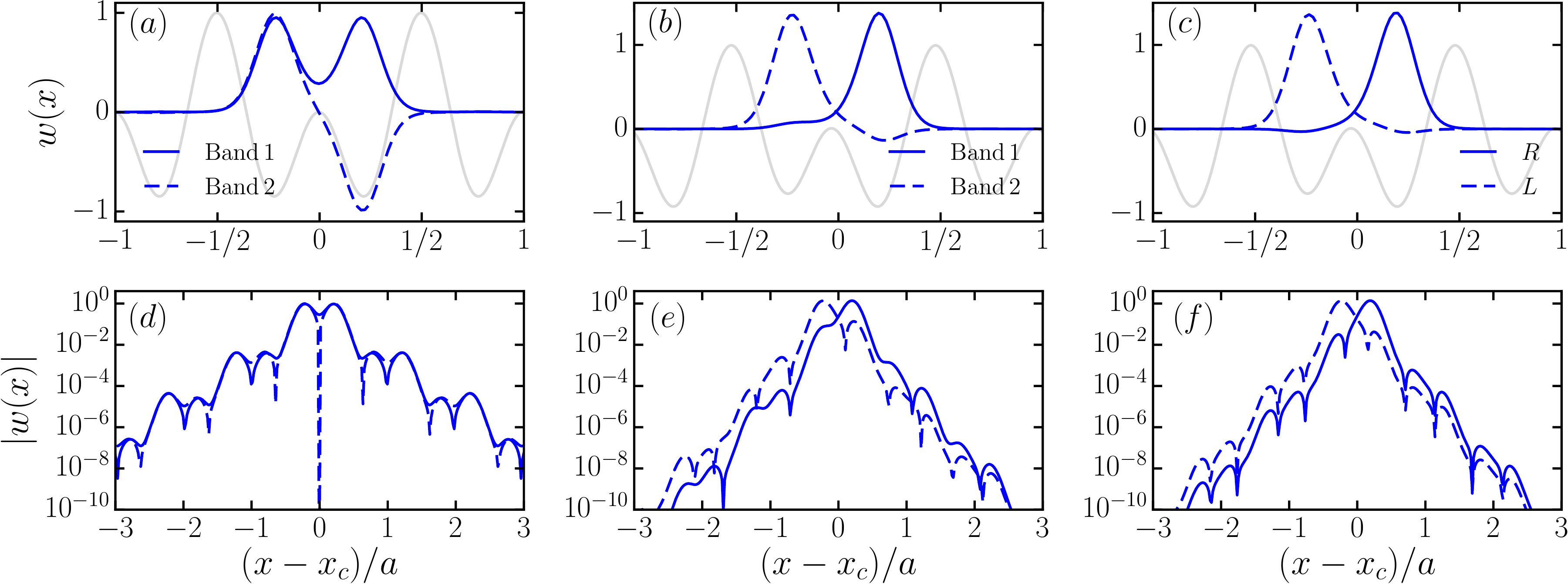}
    \caption{(color online) Plots of normalized Wannier functions $w_{j_c,\alpha}(x)$ and generalized Wannier functions $v_{j_c,\eta}(x)$ in the center of the lattice as a function of $x$. Here, wave functions and position are in units of $1/\sqrt{M_xa}$ and lattice period $a$, respectively. Panels (a) and (d) show the $\alpha=1,2$ Wannier functions for a symmetric lattice on a linear and logarithmic scale, respectively. For clarity, we have shifted the $x$ axis by $x_c$, such that the center of the interval is at the origin. Here, $M_x=21$, $N_{\rm band}=53$, $k_Lb=0.25\pi$, $V_0/E_R=35$ and $V_1/V_0=1.3$. The solid and dashed blue curves represent the $1^{\rm st}$ and $2^{\rm nd}$ band, respectively. The gray line represents the symmetric double-well potential. Panels (b) and (e) show similar plots, but now for an asymmetric lattice with $k_Lb=0.275\pi$ with other parameters unchanged. Panels (c) and (f) show the generalized Wannier functions at the $L$ and $R$ wells within a double-well for the same lattice as used in panels (b) and (e). The solid and dashed blue lines represent $v_{j_c,R}(x)$ and $v_{j_c,L}(x)$, respectively.}
    \label{fig:wfn}
  \end{figure*}

  The eigen functions $|\lambda\rangle$ with dispersion energy $\epsilon_{\lambda}$ of $H_0$ with $\lambda\in\{1,\ldots, 2N_x+1\}$ can be grouped into $N_{\rm band}$ bands containing $M_{x}$ discrete quasi-momenta. This implies that both $N_{\rm band}$ and $M_x$ must be odd as $M_xN_{\rm band}=2N_x+1$. In fact, the lowest $M_x$ eigen energies correspond to the $1^{\rm st}$ band, the next set corresponds to the $2^{\rm nd}$ band, and so on. It can be separately shown from the $\phi_n(x)$ that the allowed quasi-momenta are 
\begin{align}
  \label{eq:kdvr}
  k_x & = \frac{2p}{M_x}\frac{\pi}{a}, \quad p=0,\pm 1,\pm 2,\ldots,\pm \frac{1}{2}\left(M_x-1\right),
\end{align}
  such that $-\pi/a\le k_x \le \pi/a$ and $k_x$ stays within the $1^{\rm st}$ Brillouin zone. It is noteworthy that $N_{\rm band}$ also corresponds to the number of grid points within each unit cell. For real potentials $V(x)$, the eigen energies for $\pm k_x$ are degenerate. Consequently, the single eigen state with zero quasi-momentum can be easily located from the dispersion $\epsilon_{\lambda}$. For other quasi-momenta, we can locate the pair of real eigen functions with degenerate $\epsilon_{\lambda}$ and compute the $2\times 2$ matrix of the momentum operator. The eigen values of the momentum operator gives the quasi momentum $k_x$, thus leading to the assignment of the band dispersion $\epsilon_{\lambda}\to\epsilon_{\alpha}(k_x)$ with band index $\alpha$. (Diagonal elements of the momentum operator are strictly zero, as the eigen functions of $H_0$ are real and periodic on interval $[0,M_xa]$)

Figure~\ref{fig:DVRvsPW}(a) shows numerical results for the double-well band dispersion at $k_x=0$ for the lowest two bands using the PW basis. We find that energy differences become ``noisy'' beyond $N_{\rm PW}> 25$ basis vectors and convergence is reached with uncertainties of $2\times 10^{-12}E_R$ independent of the lattice asymmetry and band. This uncertainty should be compared with the band gap, $\approx \Delta$, between the two bands, which is on the order of $E_R$ for typical lattice depths, and is close to the numerical accuracy to be expected using double-precision arithmetic. Figure~\ref{fig:DVRvsPW}(b) shows similar data, but now obtained for the DVR calculations as a function of $N_{\rm band}$ and $M_x=3$. The integers $N_{\rm band}$ and $N_{\rm PW}$ can be directly compared as they both correspond to the number of energy bands obtained within the corresponding calculation. We find that convergence is reached for $N_{\rm band}> 25$ with uncertainties of $2\times 10^{-11}E_R$ independent of the lattice asymmetry and band. For PW calculations with $N_{PW}>25$ and DVR calculations with $N_{\rm band}>25$, the largest uncertainty is independent of quasi-momentum.

Figure~\ref{fig:DVRvsPW}(c) compares the $k_x=0$ band dispersion computed with the DVR and PW basis, respectively. It shows that for symmetric lattices, the DVR and PW results agree to within the uncertainty of the DVR calculation. For asymmetric lattices, however, they converge to different values. The inset further highlights the difference between symmetric and asymmetric lattices by studying the difference of the band dispersion as a function of lattice asymmetry $k_Lb$. We find that the value of $\epsilon_{\alpha}(k_x=0)$ for the DVR is always larger than the PW result and the difference is symmetric around $k_Lb=\pi/4$. The two results only agree infinitesimally close to $k_Lb=\pi/4$. Furthermore, we find that the discrepancy is the same independent of quasi-momentum. As we will show in Sec.~\ref{sec:validity}, this constant offset, nevertheless, leads to tunneling energies that are more accurate than  might naively be expected.

Although we have focused on DVR-based band structure calculations for a one dimensional lattice, the method can be extended to higher dimensional non-separable lattices, such as graphene. The simplest approach is based on the realization that it is always possible to construct a non-primitive unit cell with orthogonal unit vectors such that  the higher-dimensional kinetic-energy operator is separable along the unit vector directions and Eq.~\eqref{eq:dvr2} can be directly used. Alternatively, we construct DVR basis functions from plane-waves that are periodic over a multiple of the primitive lattice vectors. In this case, the kinetic energy is not separable, but can still be expressed in terms of trigonometric functions. We, however, note that for a $d$-dimensional lattice the matrix size of the single-particle Hamiltonian in the DVR method will be $M^d$ times the  size of the corresponding PW matrix, where $M$ is the number of discrete quasi-momentum points along an axis. This implies that the determination of the eigen pairs with the DVR method is computationally more intensive, but is guaranteed to lead to real-valued eigen vectors.
\end{section}

\begin{section}{DVR-based Wannier functions}
  \label{sec:wfn}
  In this section we numerically construct real-valued Wannier functions localized within unit cells and generalized Wannier functions localized near the potential minina in each double well from superpositions of our real-valued DVR eigen functions. Here, we describe a method for constructing these Wannier functions based on  Refs.~\cite{kivelson_wannier_1982,uehlinger_artificial_2013}. 

Kivelson \cite{kivelson_wannier_1982} showed that for symmetric lattices with inversion symmetry, real-valued Wannier functions for band $\alpha$ are eigen states of the projected position operator $\hat x_{\alpha} = {\cal P}_{\alpha}\,\hat x\, {\cal P}_{\alpha}$, where ${\cal P}_{\alpha}$ is the projection operator on the eigen states of band $\alpha$. The spacing between neighboring eigen values of this projected operator is a lattice constant.

    We extend this approach for constructing real-valued Wannier functions to both symmetric and asymmetric lattices lacking inversion symmetry, even though there is no formal proof that for asymmetric lattices eigen functions of the position operator are Wannier functions. We term our functions ``DVR-based'' Wannier functions. Following the previous section, the DVR eigen functions $|\lambda\rangle$ can be grouped into bands $\alpha$. In fact, we have $|\lambda\rangle=|m,\alpha\rangle$, with $m\in\{1,\ldots,M_x\}$ and projector ${\cal P}_{\alpha}=\sum_{m}|m,\alpha\rangle\langle m,\alpha|$. We construct the matrix $\langle m,\alpha|\hat x|m',\alpha\rangle$ over all $m$ and $m'$ in the same band $\alpha$ using that $\langle x_i|\hat x|x_{i'}\rangle=x_i\delta_{ii'}$ to good approximation. Diagonalization leads to real DVR-based Wannier functions $w_{j,\alpha}(x)$ for unit cell $j=\{1,\ldots,M_x\}$ and as we will show in Sec.~\ref{sec:validity}, they reproduce the tunneling energies with great accuracy.

  \begin{figure}
    \centering
    \includegraphics[width=3.2in]{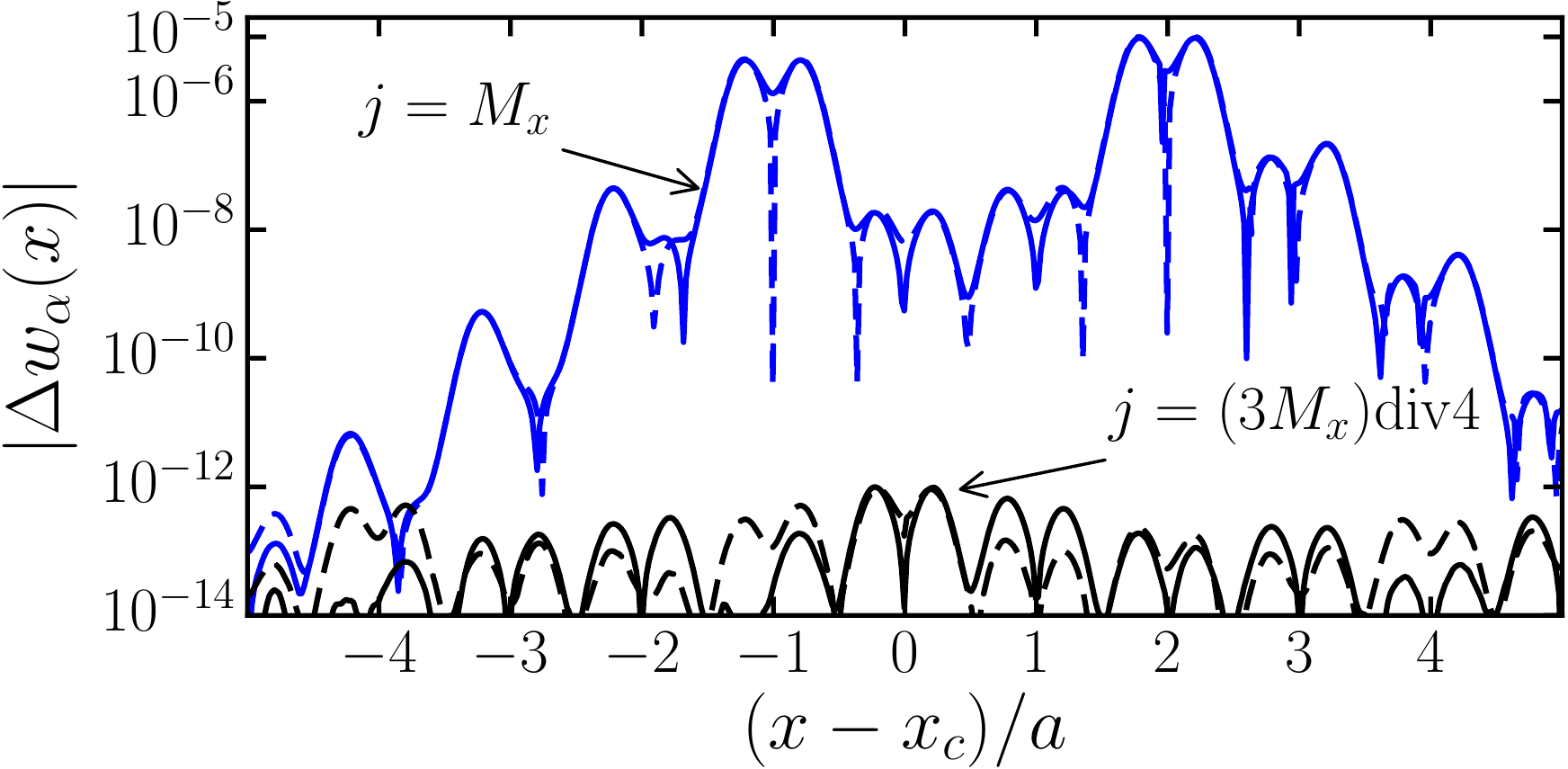}
    \caption{(color online) Graphs of difference between (shifted) Wannier functions $w_{j,\alpha}(x)$ and that at the center of the lattice. Plotted are $\Delta w_{\alpha}(x)=|w_{j,\alpha}(x-[j-j_c]a)|-|w_{j_c,\alpha}(x)|$ for unit cells $j=(3M_x){\rm div} 4$ (black curves) and $M_x$ (blue curves) as a function of $x$ in units of lattice period $a$. The argument $x-[j-j_c]a$ is computed assuming modular arithmetic on interval $M_xa$. Solid and dashed lines correspond to bands $\alpha=1$ and $2$, respectively. The plot is for a symmetric lattice with $k_Lb=0.25\pi$, $V_0=35E_R$, $V_1/V_0=1.3$, $M_x=21$ and $N_{\rm band}=53$.}
    \label{fig:wfn-compare}
  \end{figure}

  \begin{figure}
    \centering
    \includegraphics[width=3.2in]{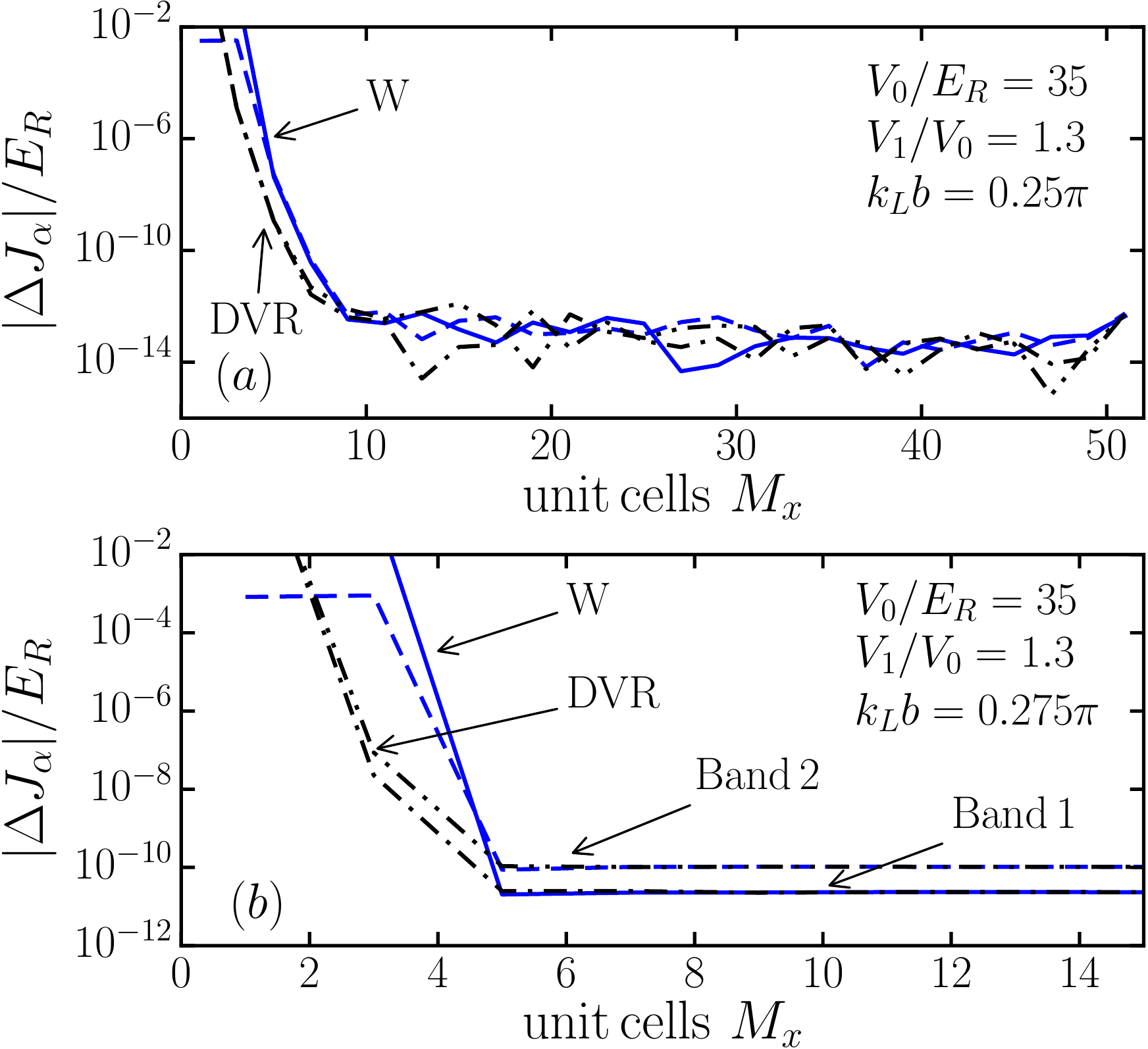}
    \caption{(color online) Comparison of the nearest neighbor tunneling energies $J_{\alpha}^{\rm PW}$, $J_{\alpha}^{\rm DVR}$ and $J_{\alpha}^{\rm W}$ for bands $\alpha=\{1,2\}$, as computed using the Fourier transform of the band dispersion from the PW and DVR calculations, and the DVR band Wannier functions $w_{\alpha}(x)$, respectively. 
      (a) Tunneling energy comparison for a symmetric lattice with $k_Lb=0.25\pi$. Plotted are $\Delta J_{\alpha}=J_{\alpha}^{\rm DVR}-J_{\alpha}^{\rm PW}$ (black curves labeled DVR) and $\Delta J_{\alpha}=J_{\alpha}^{\rm W}-J_{\alpha}^{\rm PW}$ (blue curves labeled W) in units of $E_R$ as a function of the number of unit cells $M_x$. Solid and dashed lines correspond to bands $\alpha=1$ and $2$, respectively. We used $V_0=35E_R$, $V_1/V_0=1.3$ and $N_{\rm PW}=N_{\rm band}=35$. (b) Similar plot for an asymmetric lattice with $k_Lb=0.275\pi$ with other parameters unchanged.}
    \label{fig:hopcompare1}
  \end{figure}

    Generalized Wannier functions $v_{j,\eta}(x)$ localized in the $\eta=L$ and $R$ wells of Fig.~\ref{fig:double-well} can be constructed by creating superpositions of DVR functions from multiple bands. In our case, we restrict the bands to $\alpha\in\{1,2\}$ and compute the eigen functions of the projected position operator ${\cal P} \hat x {\cal P}$, where ${\cal P}=\sum_{m,\alpha=1,2}|m,\alpha\rangle\langle m,\alpha|$. This approach is used for both symmetric and asymmetric lattices.

    Figures~\ref{fig:wfn} (a) and (d) show numerical Wannier functions $w_{j,\alpha}(x)$ for a symmetric lattice with band index $\alpha\in\{1,2\}$ on a linear and logarithmic scale, respectively. The Wannier function is localized in the unit cell at the center of the lattice with $j=j_c\equiv (M_x+1){\rm div}2$ and $x_c=M_xa/2$. For the symmetric lattice, both $w_{j_c,1}(x)$ and $w_{j_c,2}(x)$ are, however, spread over the two wells in the unit cell. Figures~\ref{fig:wfn} (b) and (e) show similar plots for an asymmetric lattice, while Figs.~\ref{fig:wfn} (c) and (f) show  generalized Wannier functions $v_{j_c,\eta}(x)$ with $\eta\in\{L,R\}$ based on the first two bands for the same lattice parameters. Owing to a large asymmetry for these last four panels, the band gap between the two lowest bands is large. We thus expect $w_{j_c,1}(x)\approx w_{j_c,R}(x)$ and $w_{j_c,2}(x)\approx w_{j_c,L}(x)$ as indeed observed when comparing Figs.~\ref{fig:wfn} (b) and (c). It is, however, interesting to note that the $v_{j_c,\eta}(x)$'s and $w_{j_c,\alpha}(x)$'s are not exactly the same. In fact, $v_{j_c,\eta}(x)$ is more localized within the $L$ and $R$ wells compared to $w_{j_c,\alpha}(x)$. For even larger asymmetries, this difference in localization persists and the ``tail'' of $w_{j_c,\alpha}(x)$ does not approach $v_{j_c,\eta}(x)$, leading to marked differences in the calculated BH parameters, as will be shown in Sec.~\ref{sec:int}.

Figure~\ref{fig:wfn-compare} shows a comparison of Wannier functions  for a symmetric lattice computed at different unit cells. We find that the difference between the Wannier functions at the edge and the center is of the order of $10^{-5}/\sqrt{M_xa}$ for all $x$. For all other unit cells, the difference from the central Wannier function is of the order of $10^{-13}/\sqrt{M_xa}$, which is close to our numerical accuracy. One of such a difference with $j=(3M_x){\rm div}4$ is shown in the figure. Hence, the shape of our DVR-based Wannier functions are mostly independent of unit cell. This observation remains true for asymmetric lattices.

\end{section}

\begin{section}{Tunneling energies based on DVR-based Wannier functions}
  \label{sec:validity}
  
  In Sec.~\ref{sec:wfn} we showed that the Wannier functions and generalized Wannier functions within a double-well can be constructed from DVR eigen vectors. In this section we use these functions to compute tunneling energies and discuss their accuracy. In particular, the accuracy of the single band Wannier functions are ascertained in Sec.~\ref{subsec:wfn} by comparing band tunneling energies as they only depend on the band dispersion and should be independent of the choice of Wannier functions. Tunneling energies between neighboring $L$ and $R$ wells are computed in Sec.~\ref{subsec:lfn} and a corresponding tight-binding (TB) model is shown to have significant contributions from tunneling energy terms between next-nearest neighbors and beyond.

  \begin{subsection}{Band tunneling energies}
    \label{subsec:wfn}
    
  \begin{figure}
    \centering
    \includegraphics[width=3.2in,height=1.65in]{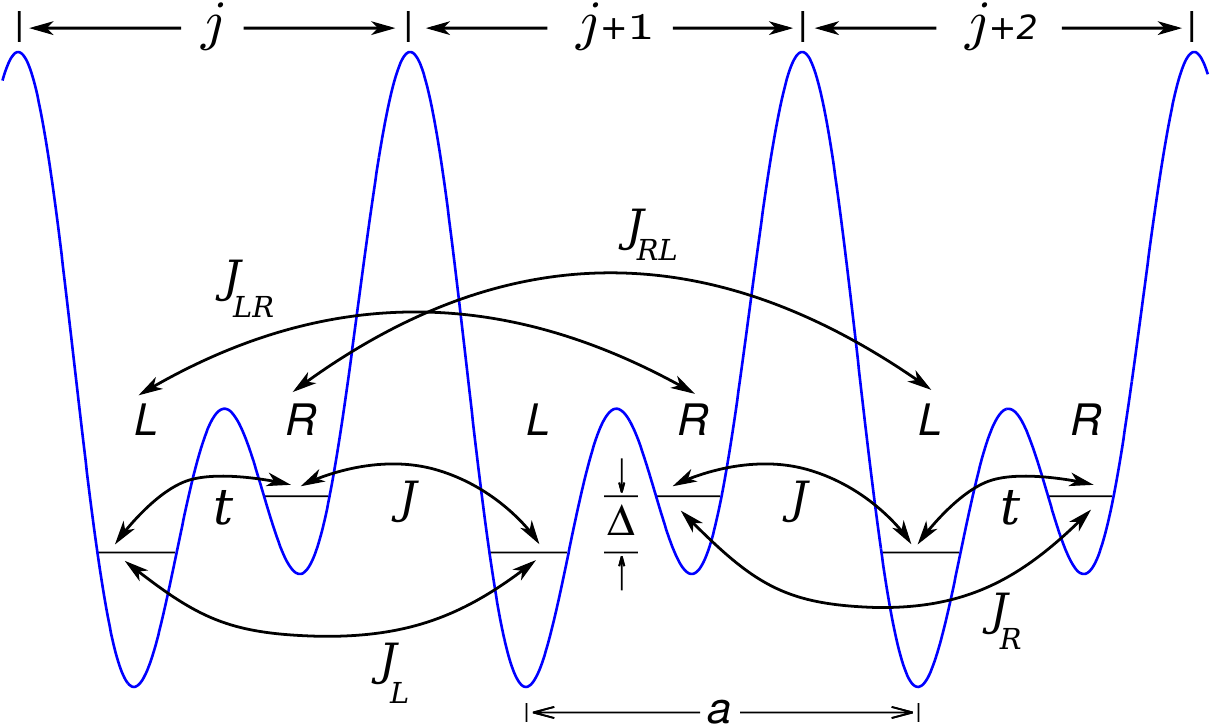}
    \caption{(color online) Tight-binding Hamiltonian based on the lowest two bands for an asymmetric double-well optical lattice. The figure shows various tunneling energies $t$, $J$, $J_L$, $J_R$, $J_{LR}$ and $J_{RL}$ between three neighboring unit cells. The energy gap between the two on-site energies is $\Delta$ and the lattice has period $a$.}
    \label{fig:tbwells}
  \end{figure}

 Tunneling energies are defined by the matrix elements $J_{\alpha}=\langle w_{j,\alpha}|H_0|w_{j',\alpha}\rangle$  over the  Wannier functions of band $\alpha$ localized in unit cells $j$ and $j'$. Here,  $H_0$ is the single-particle Hamiltonian. We mainly focus on nearest-neighbor tunneling with $j'=j\pm 1$. Formally, the $J_{\alpha}$ should only depend on $|j-j'|$.

There are three different ways to obtain tunneling energies. The first is to use our DVR-based Wannier functions for band $\alpha$ as computed in Sec.~\ref{sec:wfn} and calculate the matrix element. We label them $J_{\alpha}^{\rm W}$. The other two methods rely on the usual definition of a Wannier function as a ``Fourier transform'' of Bloch functions of the corresponding band. With this definition, the tunneling energies only depend on the band dispersion $\epsilon_{\alpha}(k_x)$ and between nearest-neighbor unit cells $(|j-j'|=1)$ is given by
\begin{align}
   \label{eq:bandhop12}
   J_{\alpha}=\frac{a}{2\pi}\int_{-\pi/a}^{\pi/a}\cos(k_xa)\epsilon_{\alpha}(k_x)dk_x,
 \end{align}
 independent of $j$. The tunneling energy can now be determined either by substituting $\epsilon_{\alpha}(k_x)$ calculated using the PW basis or by using the band dispersion obtained from the DVR method. We refer to these tunneling energies by $J_{\alpha}^{\rm PW}$ and $J_{\alpha}^{\rm DVR}$, respectively.

 Figure~\ref{fig:hopcompare1} shows a comparison between tunneling energies $J_{\alpha}^{\rm PW}$, $J_{\alpha}^{\rm DVR}$ and $J_{\alpha}^{\rm W}$ as a function of the number of unit cells. The energy $J_{\alpha}^{\rm W}$ has been computed using ``DVR-based'' Wannier function for the central unit cell. We find that for a symmetric lattice (panel (a)) convergence is reached for $M_x> 9$ unit cells, with uncertainties of $2\times 10^{-13}E_R$ for all methods. This confirms the central idea of Ref.~\cite{kivelson_wannier_1982}, that Wannier functions are eigen states of the $\hat x_{\alpha}$ operator for symmetric lattices. Figure~\ref{fig:hopcompare1} (b) shows $J_{\alpha}^{\rm DVR}-J_{\alpha}^{\rm PW}$ and $J_{\alpha}^{\rm W}-J_{\alpha}^{\rm PW}$ converge to $2\times 10^{-11}E_R$ for band $1$ and $1\times 10^{-10}E_R$ for band $2$, much above the value reached for the symmetric lattice. Within the DVR calculation, however, $J_{\alpha}^{\rm W}$ and $J_{\alpha}^{\rm DVR}$ agree much better. The discrepancy between the PW and DVR results can be attributed to the difference in the band dispersion shown in Fig.~\ref{fig:DVRvsPW}. Nevertheless, even an uncertainty of $10^{-10}E_R$ is sufficient for all practical purposes.

We have numerically ascertained that $J_{\alpha}^{\rm W}$ does not vary with the unit cell index $j$ to better than $10^{-13}E_R$ apart from the two edge unit cells consistent with our observations on the shape of Wannier functions in Fig.~\ref{fig:wfn-compare}. In fact, the difference between the tunneling energies at the central and edge unit cell is only $10^{-8}E_R$. Consequently, the value of $J_{\alpha}^{\rm W}$ obtained from the central unit cell is better than that from the edge unit cells and agrees better with $J_{\alpha}^{\rm PW}$. In other words, a comparison with the tunneling energies $J_{\alpha}^{\rm PW}$ gives a good estimate of the accuracy of our real-valued Wannier functions.

We have also determined the next-nearest neighbor tunneling energies. For typical lattice depths, its value is two orders of magnitude or more lower than that of nearest neighbors. Its uncertainty in units of $E_R$ is the same as for nearest-neighbor tunneling energies. Hence, we conclude that the  DVR-based Wannier functions can be used to compute tunneling energies between distant neighbors.
\end{subsection}

\begin{subsection}{Tight binding tunneling energies}
  \label{subsec:lfn}

  \begin{figure}
    \centering
    \includegraphics[width=3.2in]{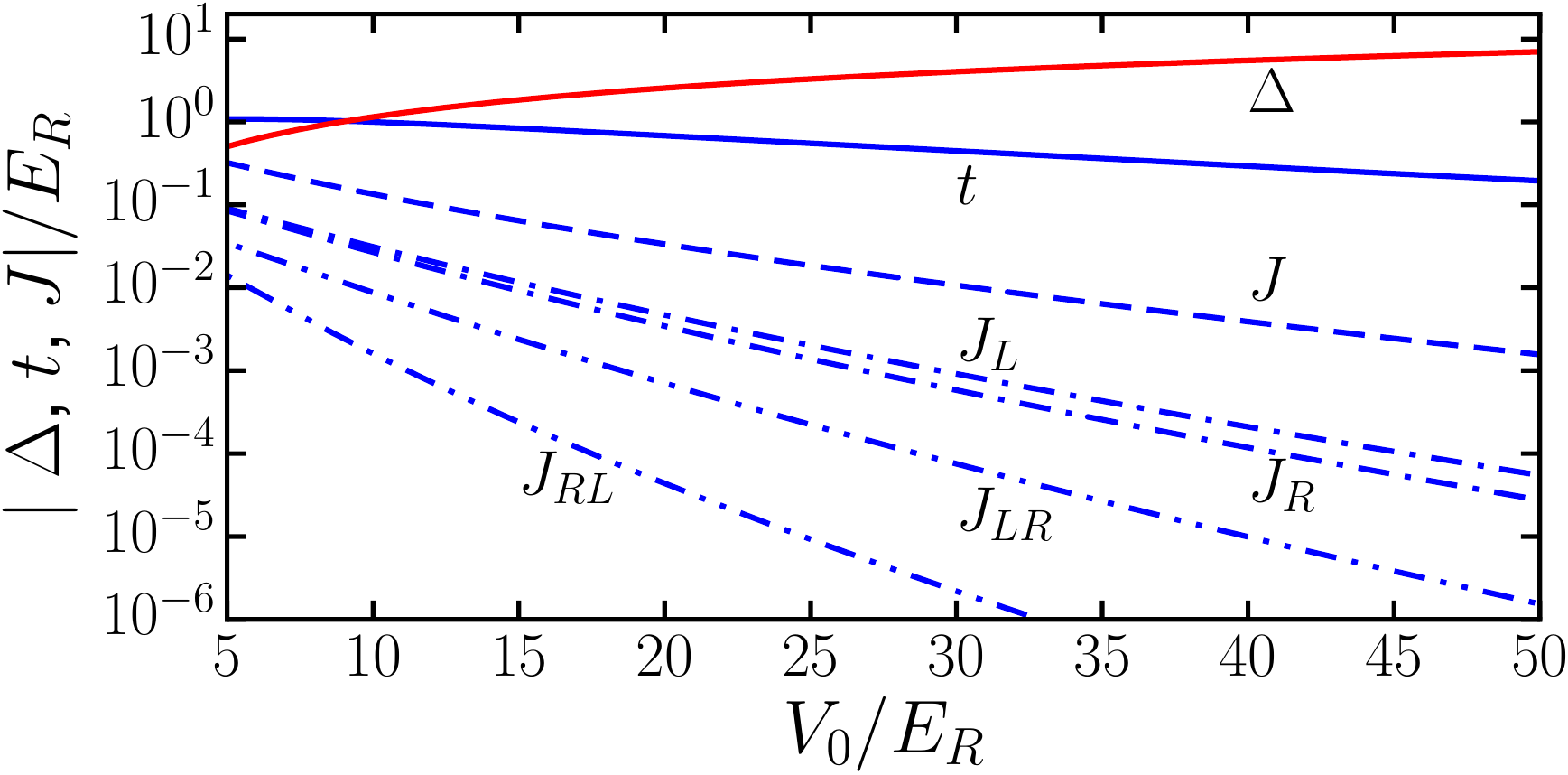}
    \caption{(color online) Log-linear plot of the absolute value of tunneling energies $t$, $J$, $J_L$, $J_R$, $J_{LR}$, $J_{RL}$ and energy-gap $\Delta$ in units of $E_R$ as a function of lattice depth $V_0$. The plot is for an asymmetric lattice with $k_Lb=0.275\pi$ and $V_1/V_0=1.3$.}
    \label{fig:PWTB}
  \end{figure}

It is often useful to write down a tight-binding Hamiltonian in terms of $L$ and $R$ wells defined in Fig.~\ref{fig:double-well} and with hopping parameters computed from our generalized Wannier functions with the lowest on-site energies $\langle v_{j,\eta}|H_0|v_{j',\eta}\rangle$. Figure~\ref{fig:tbwells} defines tunneling energies between adjacent unit cells and the energy gap $\Delta$ between the on-site energies based on the lowest two bands of our $H_0$. The largest parameters are given by $t=\langle w_{j,L}|H_0|w_{j,R}\rangle$  and $J=\langle w_{j,R}|H_0|w_{j+1,L}\rangle$, where $j$ is the unit cell index. Similar expressions can be written down for other parameters. The value of these tunneling energies depends on the definition of the generalized Wannier functions and cannot be extracted from a transformation of the band dispersion energies. Finally, we note that all coefficients are real-valued.

 Figure~\ref{fig:PWTB} shows the largest seven hopping parameters of our TB model as a function of lattice depth $V_0$ for an asymmetric lattice. As expected, the tunneling energies decrease with lattice depth, while simultaneously $\Delta$ increases. For fixed lattice depth the tunneling energies are smaller the further the atom has to hop.

 The TB Hamiltonian for two modes within a unit cell can be diagonalized analytically by a  transformation to quasi-momentum space. In fact, the eigen energies are 
\begin{align}
  \label{eq:tbbands}
  \epsilon_{\alpha}^{\rm TB}(k_x)&=-(J_R+J_L)\cos{k_xa}\\
  &\quad\mp\sqrt{\left[(J_R-J_L)\cos{k_xa}-\Delta/2\right]^2+|f(k_x)|^2}\nonumber,
\end{align}
 where $\mp$ correspond to  bands $\alpha=1$ and $2$, respectively, and $f(k_x)=t+Je^{-ik_xa}+J_{LR}e^{ik_xa}+J_{RL}e^{-2ik_xa}$.
The band tunneling energies $J_{\alpha}^{\rm TB}$ can be obtained by substituting $\epsilon_{\alpha}^{\rm TB}(k_x)$ into Eq.~\eqref{eq:bandhop12} and performing the Fourier transform. 

We can now compare the band tunneling energies of our TB simulations with those of the exact band structure calculations using the PW basis. We find that the difference between the TB and PW result is within approximately $50\%$ for both bands when we only include nearest-neighbor tunneling energies $t$ and $J$ and $5\%$ when in addition next nearest-neighbor tunneling energies $J_L$ and $J_R$ are included, and this stays nearly the same upon including the next to next-nearest neighbor hopping terms $J_{LR}$ and $J_{RL}$. These differences are almost independent of the lattice depth and consistent with results of Ref.~\cite{modugno_maximally_2012} who based their calculations on complex-valued maximally-localized Wannier functions. The TB result can get better if we include more tunneling energies and allow atoms to hop even further.
  \end{subsection}
  
\end{section}

\begin{section}{Interaction energies}
  \label{sec:int}

  \begin{figure}
    \centering
    \includegraphics[width=3.2in]{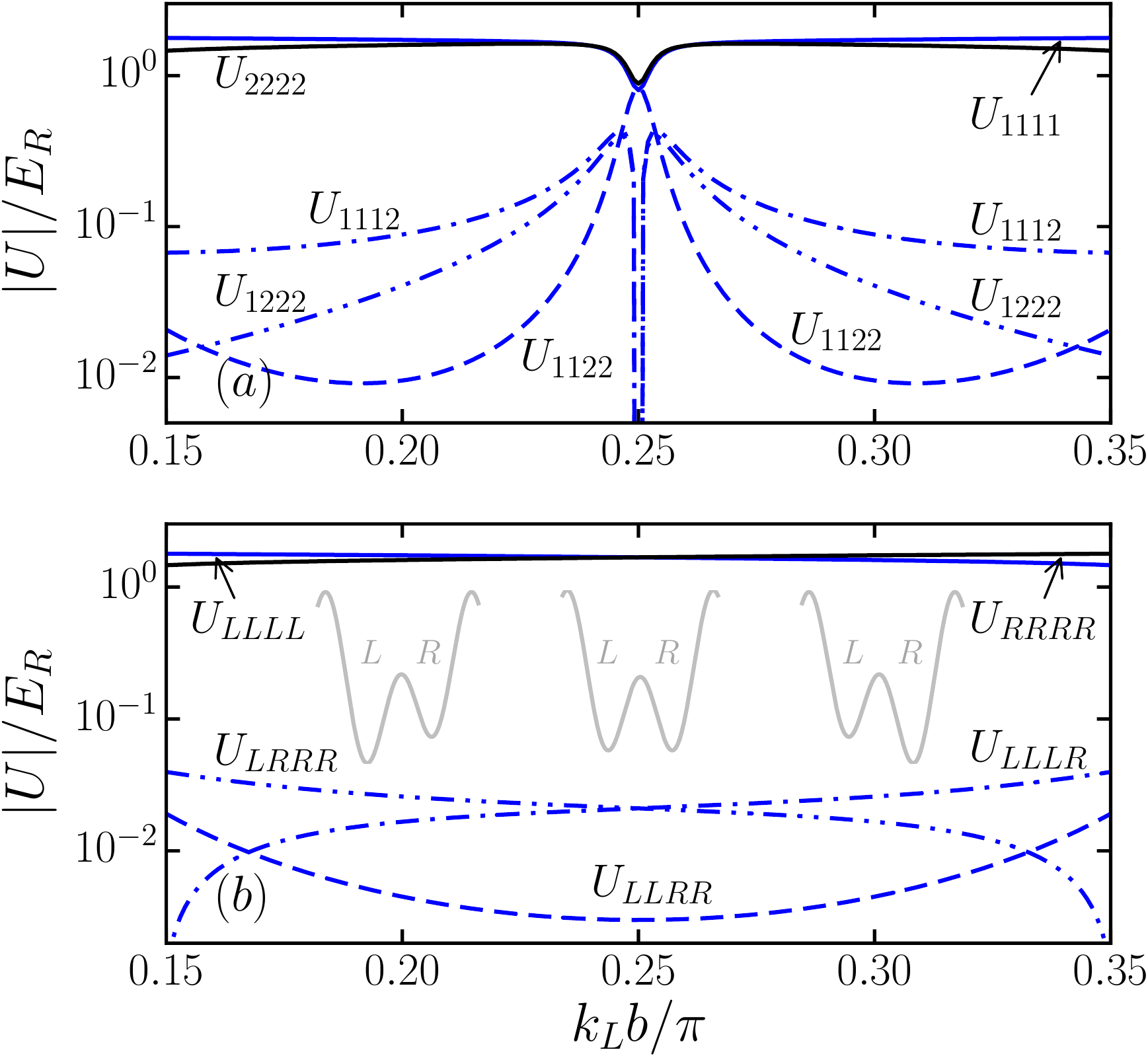}
    \caption{(color online) Two-body interaction energies in the Hubbard model for ${}^{\rm 87}$Rb in units of $E_R$ as a function of lattice asymmetry $k_Lb/\pi$. We use $V_0/E_R=35$, $V_1/V_0=1.3$, $V_2/E_R=70$ and scattering length $a_s=5.3$ nm. Panel (a) shows interaction energies $U_{\alpha_1\alpha_2\alpha_3\alpha_4}$ in the band basis with $\alpha\in\{1,2\}$. Panel (b) shows interaction energies $U_{\eta_1\eta_2\eta_3\eta_4}$ in the $LR$ basis with $\eta\in\{L,R\}$. From left to right the insets  show a schematic of a double-well potential for lattice asymmetries $k_Lb<0.25\pi$, $k_Lb=0.25\pi$ and $k_Lb>0.25\pi$, respectively.}
    \label{fig:Uvsb}
  \end{figure}

  We have shown the excellent accuracy of the DVR-based Wannier functions in Sec.~\ref{sec:validity}. In this section, we use these functions to study the two-body atom-atom interaction terms in the Hubbard model. So far, we have focused on the double-well lattice along the $x$ axis. We can extend the calculations to the perpendicular directions and obtain the corresponding Wannier functions. Owing to the large band gap between the $1^{\rm st}$ and $2^{\rm nd}$ bands along perpendicular directions compared to that along the $x$ direction, only the ground band is considered. Thus, the full three-dimensional band Wannier functions are $w_{\alpha}(\vec x)=w_{j_c,\alpha}(x)w(y)w(z)$, where band index $\alpha\in\{1,2\}$ and $w(y)$, $w(z)$ are the ground-band Wannier functions at the center of the lattice along the perpendicular directions. We note that the $y$ and $z$ Wannier functions have the same functional form as for simplicity we have assumed the same lattice depth along the perpendicular directions. Similarly, the generalized Wannier functions are $v_{\eta}(\vec x)=v_{j_c,\eta}(x)w(y)w(z)$, where $\eta\in\{L,R\}$. 

In the Hubbard model and band basis, the two-body on-site interaction energies are
  \begin{align}
    \label{eq:U12}
    U_{\alpha_1\alpha_2\alpha_3\alpha_4} = g\int w_{\alpha_1}(\vec x)w_{\alpha_2}(\vec x)w_{\alpha_3}(\vec x)w_{\alpha_4}(\vec x)d\vec x,
  \end{align}
  where $g=4\pi\hbar^2a_s/m_a$, $a_s$ is the $s$-wave scattering length and we use that the $w(\vec x)$ are real. There are five distinct coefficients: $U_{1111}$, $U_{1112}$, $U_{1122}$, $U_{1222}$ and $U_{2222}$. On-site interaction energies $U_{\eta_1\eta_2\eta_3\eta_4}$ in the $LR$ basis using the generalized Wannier functions $v_{\eta}(\vec x)$ can be similarly defined. The five distinct interactions coefficients are $U_{LLLL}$, $U_{LLLR}$, $U_{LLRR}$, $U_{LRRR}$ and $U_{RRRR}$. 

  Figures~\ref{fig:Uvsb} (a) and (b) show the two-body interaction energies $U_{\alpha_1\alpha_2\alpha_3\alpha_4}$ and $U_{\eta_1\eta_2\eta_3\eta_4}$, respectively, for ${}^{\rm 87}$Rb with $a_s=5.3$ nm as a function of the lattice asymmetry $b$, with other lattice parameters kept fixed. Figure~\ref{fig:Uvsb} (a) is symmetric around $k_Lb=\pi/4$. At the symmetry point $k_Lb=\pi/4$, $U_{1111}\lesssim U_{1122}\lesssim U_{2222}$, while $U_{1112}$ and $U_{1222}$ are strictly zero due to parity. As the lattice becomes asymmetric, $U_{1111}$ and $U_{2222}$ double their strength, $U_{1122}$ rapidly decreases, while $U_{1112}$ and $U_{1222}$ have a maximum but remain relatively large. 

  Figure~\ref{fig:Uvsb} (b) shows that the $U_{\eta_1\eta_2\eta_3\eta_4}$ have a much smoother dependence on the asymmetry than the $U_{\alpha_1\alpha_2\alpha_3\alpha_4}$. For all asymmetries, we observe that $U_{LLLL}$ and $U_{RRRR}$ are much larger than the other energies. Moreover, $U_{LLLL}=U_{RRRR}$ for a symmetric lattice, and $U_{RRRR}>U_{LLLL}$ for $k_Lb>\pi/4$. This behavior is reversed for $k_Lb<\pi/4$. This is a consequence of the fact that $v_R(\vec x)$ is more confined than $v_L(\vec x)$ for $k_Lb>\pi/4$ and vice versa. Interestingly, the density-induced tunneling energies $U_{LRRR}$ and $U_{LLLR}$  are, in general, larger than the density-density term $U_{LLRR}$. The former coefficients lead to terms in a Hubbard model where an atom hops from one well to the other in a unit cell, while the latter coefficient leads to either a long-range density-density interaction or pair hopping. The relative size of these energies highlight the limits of Hubbard models that do or do not include specific two-body terms \cite{jurgensen_density-induced_2012}.

The two-body interaction energies within the two bases can be compared in several limits of the lattice asymmetry. For $k_Lb>\pi/4$ and very large asymmetries where $U_{1122}\ll U_{1111}$ the Wannier functions $w_{1}(\vec x)$ approach $v_{R}(\vec x)$ (similarly, $w_2(\vec x)$ approaches $v_L(\vec x)$) and, thus, $U_{1111}\to U_{RRRR}$ and $U_{2222}\to U_{LLLL}$. In fact, for as low as $k_Lb=0.26\pi$, $U_{1111}\approx 0.95\,U_{RRRR}$. On the other hand, for a symmetric lattice we can write $w_{1,2}(\vec x)\approx (v_L(\vec x)\pm v_R(\vec x))/\sqrt{2}$ and $U_{LLLL}=U_{RRRR}$, which leads to $U_{1111}$, $U_{1122}$, $U_{2222}\approx U_{LLLL}/2$. The additional realization that $U_{LLLL}$ and $U_{RRRR}$ are nearly insensitive to asymmetry also explains the doubling in value of $U_{1111}$ and $U_{2222}$ near $k_Lb=\pi/4$. 

Even though the Wannier functions $w_{1}(\vec x)$ and $v_{R}(\vec x)$ approach each other for large asymmetries and $k_Lb>\pi/4$, the function $v_{R}(x)$ is always more confined than $w_{1}(x)$. Consequently, cross-terms $U_{1122}$, $U_{1112}$ and $U_{1222}$, which depend on the tail of the Wannier functions, are always larger than the corresponding cross-terms $U_{LLRR}$, $U_{LLLR}$ and $U_{LRRR}$. 
\end{section}

\begin{section}{conclusion}
  \label{sec:conclude}
  We have shown that real-valued Wannier functions can be efficiently constructed for both symmetric and asymmetric periodic potentials or optical lattices. The first step involves obtaining the single-particle band structure and real-valued eigen vectors using a Discrete Variable Representation (DVR). A Fourier grid DVR based on basis functions with periodic boundary conditions is shown to have excellent numerical accuracy compared to a direct calculation based on plane-waves. In the next step, restricted to eigen vectors within the subspace of band $\alpha$, Wannier functions $w_{\alpha}(x)$ localized within a unit cell are obtained as eigen states of the position operator. By using eigen vectors corresponding to the lowest two bands, generalized Wannier functions $w_{\eta}(x)$ localized to $L$ and $R$ wells within a double-well can also be constructed. By a comparison of the tunneling energies, the Wannier functions are shown to reproduce the Hubbard parameters with excellent accuracy. Tunneling energies are subsequently obtained between the $L$ and $R$ wells using the generalized Wannier functions, and limits of a tight-binding containing only nearest-neighbor tunneling energies are discussed. Finally, we use these functions to study the two-body interaction energies in the BH model and discuss the relative importance of the various interaction energy terms. The numerical methods developed are general and can be applied to a wide array of optical lattice geometries in one, two or three dimensions.
\end{section}

\begin{section}{Acknowledgements}
This work has been supported by the National Science Foundation Grant No. PHY-1506343.  
\end{section}

\bibliography{references}

\begin{thebibliography}{35}
\expandafter\ifx\csname natexlab\endcsname\relax\def\natexlab#1{#1}\fi
\expandafter\ifx\csname bibnamefont\endcsname\relax
  \def\bibnamefont#1{#1}\fi
\expandafter\ifx\csname bibfnamefont\endcsname\relax
  \def\bibfnamefont#1{#1}\fi
\expandafter\ifx\csname citenamefont\endcsname\relax
  \def\citenamefont#1{#1}\fi
\expandafter\ifx\csname url\endcsname\relax
  \def\url#1{\texttt{#1}}\fi
\expandafter\ifx\csname urlprefix\endcsname\relax\def\urlprefix{URL }\fi
\providecommand{\bibinfo}[2]{#2}
\providecommand{\eprint}[2][]{\url{#2}}

\bibitem[{\citenamefont{Kivelson}(1982)}]{kivelson_wannier_1982}
\bibinfo{author}{\bibfnamefont{S.}~\bibnamefont{Kivelson}},
  \bibinfo{journal}{Physical Review B} \textbf{\bibinfo{volume}{26}},
  \bibinfo{pages}{4269} (\bibinfo{year}{1982}).

\bibitem[{\citenamefont{Bloch et~al.}(2008)\citenamefont{Bloch, Dalibard, and
  Zwerger}}]{bloch_many-body_2008}
\bibinfo{author}{\bibfnamefont{I.}~\bibnamefont{Bloch}},
  \bibinfo{author}{\bibfnamefont{J.}~\bibnamefont{Dalibard}}, \bibnamefont{and}
  \bibinfo{author}{\bibfnamefont{W.}~\bibnamefont{Zwerger}},
  \bibinfo{journal}{Reviews of Modern Physics} \textbf{\bibinfo{volume}{80}},
  \bibinfo{pages}{885} (\bibinfo{year}{2008}).

\bibitem[{\citenamefont{Jaksch and Zoller}(2005)}]{jaksch_cold_2005}
\bibinfo{author}{\bibfnamefont{D.}~\bibnamefont{Jaksch}} \bibnamefont{and}
  \bibinfo{author}{\bibfnamefont{P.}~\bibnamefont{Zoller}},
  \bibinfo{journal}{Annals of Physics} \textbf{\bibinfo{volume}{315}},
  \bibinfo{pages}{52} (\bibinfo{year}{2005}).

\bibitem[{\citenamefont{Jaksch et~al.}(1998)\citenamefont{Jaksch, Bruder,
  Cirac, Gardiner, and Zoller}}]{jaksch_cold_1998}
\bibinfo{author}{\bibfnamefont{D.}~\bibnamefont{Jaksch}},
  \bibinfo{author}{\bibfnamefont{C.}~\bibnamefont{Bruder}},
  \bibinfo{author}{\bibfnamefont{J.~I.} \bibnamefont{Cirac}},
  \bibinfo{author}{\bibfnamefont{C.~W.} \bibnamefont{Gardiner}},
  \bibnamefont{and} \bibinfo{author}{\bibfnamefont{P.}~\bibnamefont{Zoller}},
  \bibinfo{journal}{Physical Review Letters} \textbf{\bibinfo{volume}{81}},
  \bibinfo{pages}{3108} (\bibinfo{year}{1998}).

\bibitem[{\citenamefont{Greiner et~al.}(2002)\citenamefont{Greiner, Mandel,
  Esslinger, Hänsch, and Bloch}}]{greiner_quantum_2002}
\bibinfo{author}{\bibfnamefont{M.}~\bibnamefont{Greiner}},
  \bibinfo{author}{\bibfnamefont{O.}~\bibnamefont{Mandel}},
  \bibinfo{author}{\bibfnamefont{T.}~\bibnamefont{Esslinger}},
  \bibinfo{author}{\bibfnamefont{T.~W.} \bibnamefont{Hänsch}},
  \bibnamefont{and} \bibinfo{author}{\bibfnamefont{I.}~\bibnamefont{Bloch}},
  \bibinfo{journal}{Nature} \textbf{\bibinfo{volume}{415}}, \bibinfo{pages}{39}
  (\bibinfo{year}{2002}).

\bibitem[{\citenamefont{Sebby-Strabley
  et~al.}(2006)\citenamefont{Sebby-Strabley, Anderlini, Jessen, and
  Porto}}]{sebby-strabley_lattice_2006}
\bibinfo{author}{\bibfnamefont{J.}~\bibnamefont{Sebby-Strabley}},
  \bibinfo{author}{\bibfnamefont{M.}~\bibnamefont{Anderlini}},
  \bibinfo{author}{\bibfnamefont{P.~S.} \bibnamefont{Jessen}},
  \bibnamefont{and} \bibinfo{author}{\bibfnamefont{J.~V.} \bibnamefont{Porto}},
  \bibinfo{journal}{Physical Review A} \textbf{\bibinfo{volume}{73}},
  \bibinfo{pages}{033605} (\bibinfo{year}{2006}).

\bibitem[{\citenamefont{Lee et~al.}(2007)\citenamefont{Lee, Anderlini, Brown,
  Sebby-Strabley, Phillips, and Porto}}]{lee_sublattice_2007}
\bibinfo{author}{\bibfnamefont{P.~J.} \bibnamefont{Lee}},
  \bibinfo{author}{\bibfnamefont{M.}~\bibnamefont{Anderlini}},
  \bibinfo{author}{\bibfnamefont{B.~L.} \bibnamefont{Brown}},
  \bibinfo{author}{\bibfnamefont{J.}~\bibnamefont{Sebby-Strabley}},
  \bibinfo{author}{\bibfnamefont{W.~D.} \bibnamefont{Phillips}},
  \bibnamefont{and} \bibinfo{author}{\bibfnamefont{J.~V.} \bibnamefont{Porto}},
  \bibinfo{journal}{Physical Review Letters} \textbf{\bibinfo{volume}{99}},
  \bibinfo{pages}{020402} (\bibinfo{year}{2007}).

\bibitem[{\citenamefont{Trotzky et~al.}(2008)\citenamefont{Trotzky, Cheinet,
  Fölling, Feld, Schnorrberger, Rey, Polkovnikov, Demler, Lukin, and
  Bloch}}]{trotzky_time-resolved_2008}
\bibinfo{author}{\bibfnamefont{S.}~\bibnamefont{Trotzky}},
  \bibinfo{author}{\bibfnamefont{P.}~\bibnamefont{Cheinet}},
  \bibinfo{author}{\bibfnamefont{S.}~\bibnamefont{Fölling}},
  \bibinfo{author}{\bibfnamefont{M.}~\bibnamefont{Feld}},
  \bibinfo{author}{\bibfnamefont{U.}~\bibnamefont{Schnorrberger}},
  \bibinfo{author}{\bibfnamefont{A.~M.} \bibnamefont{Rey}},
  \bibinfo{author}{\bibfnamefont{A.}~\bibnamefont{Polkovnikov}},
  \bibinfo{author}{\bibfnamefont{E.~A.} \bibnamefont{Demler}},
  \bibinfo{author}{\bibfnamefont{M.~D.} \bibnamefont{Lukin}}, \bibnamefont{and}
  \bibinfo{author}{\bibfnamefont{I.}~\bibnamefont{Bloch}},
  \bibinfo{journal}{Science} \textbf{\bibinfo{volume}{319}},
  \bibinfo{pages}{295} (\bibinfo{year}{2008}).

\bibitem[{\citenamefont{Atala et~al.}(2013)\citenamefont{Atala, Aidelsburger,
  Barreiro, Abanin, Kitagawa, Demler, and Bloch}}]{atala_direct_2013}
\bibinfo{author}{\bibfnamefont{M.}~\bibnamefont{Atala}},
  \bibinfo{author}{\bibfnamefont{M.}~\bibnamefont{Aidelsburger}},
  \bibinfo{author}{\bibfnamefont{J.~T.} \bibnamefont{Barreiro}},
  \bibinfo{author}{\bibfnamefont{D.}~\bibnamefont{Abanin}},
  \bibinfo{author}{\bibfnamefont{T.}~\bibnamefont{Kitagawa}},
  \bibinfo{author}{\bibfnamefont{E.}~\bibnamefont{Demler}}, \bibnamefont{and}
  \bibinfo{author}{\bibfnamefont{I.}~\bibnamefont{Bloch}},
  \bibinfo{journal}{Nature Physics} \textbf{\bibinfo{volume}{9}},
  \bibinfo{pages}{795} (\bibinfo{year}{2013}).

\bibitem[{\citenamefont{Tarruell et~al.}(2012)\citenamefont{Tarruell, Greif,
  Uehlinger, Jotzu, and Esslinger}}]{tarruell_creating_2012}
\bibinfo{author}{\bibfnamefont{L.}~\bibnamefont{Tarruell}},
  \bibinfo{author}{\bibfnamefont{D.}~\bibnamefont{Greif}},
  \bibinfo{author}{\bibfnamefont{T.}~\bibnamefont{Uehlinger}},
  \bibinfo{author}{\bibfnamefont{G.}~\bibnamefont{Jotzu}}, \bibnamefont{and}
  \bibinfo{author}{\bibfnamefont{T.}~\bibnamefont{Esslinger}},
  \bibinfo{journal}{Nature} \textbf{\bibinfo{volume}{483}},
  \bibinfo{pages}{302} (\bibinfo{year}{2012}).

\bibitem[{\citenamefont{Jo et~al.}(2012)\citenamefont{Jo, Guzman, Thomas,
  Hosur, Vishwanath, and Stamper-Kurn}}]{jo_ultracold_2012}
\bibinfo{author}{\bibfnamefont{G.-B.} \bibnamefont{Jo}},
  \bibinfo{author}{\bibfnamefont{J.}~\bibnamefont{Guzman}},
  \bibinfo{author}{\bibfnamefont{C.~K.} \bibnamefont{Thomas}},
  \bibinfo{author}{\bibfnamefont{P.}~\bibnamefont{Hosur}},
  \bibinfo{author}{\bibfnamefont{A.}~\bibnamefont{Vishwanath}},
  \bibnamefont{and} \bibinfo{author}{\bibfnamefont{D.~M.}
  \bibnamefont{Stamper-Kurn}}, \bibinfo{journal}{Physical Review Letters}
  \textbf{\bibinfo{volume}{108}}, \bibinfo{pages}{045305}
  (\bibinfo{year}{2012}).

\bibitem[{\citenamefont{Lee et~al.}(2009)\citenamefont{Lee, Gr\'emaud, Han,
  Englert, and Miniatura}}]{lee_ultracold_2009}
\bibinfo{author}{\bibfnamefont{K.~L.} \bibnamefont{Lee}},
  \bibinfo{author}{\bibfnamefont{B.}~\bibnamefont{Gr\'emaud}},
  \bibinfo{author}{\bibfnamefont{R.}~\bibnamefont{Han}},
  \bibinfo{author}{\bibfnamefont{B.-G.} \bibnamefont{Englert}},
  \bibnamefont{and}
  \bibinfo{author}{\bibfnamefont{C.}~\bibnamefont{Miniatura}},
  \bibinfo{journal}{Physical Review A} \textbf{\bibinfo{volume}{80}},
  \bibinfo{pages}{043411} (\bibinfo{year}{2009}).

\bibitem[{\citenamefont{Uehlinger et~al.}(2013)\citenamefont{Uehlinger, Jotzu,
  Messer, Greif, Hofstetter, Bissbort, and
  Esslinger}}]{uehlinger_artificial_2013}
\bibinfo{author}{\bibfnamefont{T.}~\bibnamefont{Uehlinger}},
  \bibinfo{author}{\bibfnamefont{G.}~\bibnamefont{Jotzu}},
  \bibinfo{author}{\bibfnamefont{M.}~\bibnamefont{Messer}},
  \bibinfo{author}{\bibfnamefont{D.}~\bibnamefont{Greif}},
  \bibinfo{author}{\bibfnamefont{W.}~\bibnamefont{Hofstetter}},
  \bibinfo{author}{\bibfnamefont{U.}~\bibnamefont{Bissbort}}, \bibnamefont{and}
  \bibinfo{author}{\bibfnamefont{T.}~\bibnamefont{Esslinger}},
  \bibinfo{journal}{Phys. Rev. Lett.} \textbf{\bibinfo{volume}{111}},
  \bibinfo{pages}{185307} (\bibinfo{year}{2013}).

\bibitem[{\citenamefont{J\"urgensen et~al.}(2012)\citenamefont{J\"urgensen,
  Sengstock, and L\"uhmann}}]{jurgensen_density-induced_2012}
\bibinfo{author}{\bibfnamefont{O.}~\bibnamefont{J\"urgensen}},
  \bibinfo{author}{\bibfnamefont{K.}~\bibnamefont{Sengstock}},
  \bibnamefont{and} \bibinfo{author}{\bibfnamefont{D.-S.}
  \bibnamefont{L\"uhmann}}, \bibinfo{journal}{Physical Review A}
  \textbf{\bibinfo{volume}{86}}, \bibinfo{pages}{043623}
  (\bibinfo{year}{2012}).

\bibitem[{\citenamefont{L\"uhmann et~al.}(2012)\citenamefont{L\"uhmann,
  J\"urgensen, and Sengstock}}]{luhmann_multi-orbital_2012}
\bibinfo{author}{\bibfnamefont{D.-S.} \bibnamefont{L\"uhmann}},
  \bibinfo{author}{\bibfnamefont{O.}~\bibnamefont{J\"urgensen}},
  \bibnamefont{and}
  \bibinfo{author}{\bibfnamefont{K.}~\bibnamefont{Sengstock}},
  \bibinfo{journal}{New Journal of Physics} \textbf{\bibinfo{volume}{14}},
  \bibinfo{pages}{033021} (\bibinfo{year}{2012}).

\bibitem[{\citenamefont{Bissbort et~al.}(2012)\citenamefont{Bissbort,
  Deuretzbacher, and Hofstetter}}]{bissbort_effective_2012}
\bibinfo{author}{\bibfnamefont{U.}~\bibnamefont{Bissbort}},
  \bibinfo{author}{\bibfnamefont{F.}~\bibnamefont{Deuretzbacher}},
  \bibnamefont{and}
  \bibinfo{author}{\bibfnamefont{W.}~\bibnamefont{Hofstetter}},
  \bibinfo{journal}{Physical Review A} \textbf{\bibinfo{volume}{86}},
  \bibinfo{pages}{023617} (\bibinfo{year}{2012}).

\bibitem[{\citenamefont{Paul and Tiesinga}(2015)}]{paul_large_2015}
\bibinfo{author}{\bibfnamefont{S.}~\bibnamefont{Paul}} \bibnamefont{and}
  \bibinfo{author}{\bibfnamefont{E.}~\bibnamefont{Tiesinga}},
  \bibinfo{journal}{Physical Review A} \textbf{\bibinfo{volume}{92}},
  \bibinfo{pages}{023602} (\bibinfo{year}{2015}).

\bibitem[{\citenamefont{Kohn}(1959)}]{kohn_analytic_1959}
\bibinfo{author}{\bibfnamefont{W.}~\bibnamefont{Kohn}},
  \bibinfo{journal}{Physical Review} \textbf{\bibinfo{volume}{115}},
  \bibinfo{pages}{809} (\bibinfo{year}{1959}).

\bibitem[{\citenamefont{Wannier}(1962)}]{wannier_dynamics_1962}
\bibinfo{author}{\bibfnamefont{G.~H.} \bibnamefont{Wannier}},
  \bibinfo{journal}{Reviews of Modern Physics} \textbf{\bibinfo{volume}{34}},
  \bibinfo{pages}{645} (\bibinfo{year}{1962}).

\bibitem[{\citenamefont{Paul and Tiesinga}(2013)}]{paul_formation_2013}
\bibinfo{author}{\bibfnamefont{S.}~\bibnamefont{Paul}} \bibnamefont{and}
  \bibinfo{author}{\bibfnamefont{E.}~\bibnamefont{Tiesinga}},
  \bibinfo{journal}{Physical Review A} \textbf{\bibinfo{volume}{88}},
  \bibinfo{pages}{033615} (\bibinfo{year}{2013}).

\bibitem[{\citenamefont{Marzari and Vanderbilt}(1997)}]{marzari_maximally_1997}
\bibinfo{author}{\bibfnamefont{N.}~\bibnamefont{Marzari}} \bibnamefont{and}
  \bibinfo{author}{\bibfnamefont{D.}~\bibnamefont{Vanderbilt}},
  \bibinfo{journal}{Physical Review B} \textbf{\bibinfo{volume}{56}},
  \bibinfo{pages}{12847} (\bibinfo{year}{1997}).

\bibitem[{\citenamefont{Marzari et~al.}(2012)\citenamefont{Marzari, Mostofi,
  Yates, Souza, and Vanderbilt}}]{marzari_maximally_2012}
\bibinfo{author}{\bibfnamefont{N.}~\bibnamefont{Marzari}},
  \bibinfo{author}{\bibfnamefont{A.~A.} \bibnamefont{Mostofi}},
  \bibinfo{author}{\bibfnamefont{J.~R.} \bibnamefont{Yates}},
  \bibinfo{author}{\bibfnamefont{I.}~\bibnamefont{Souza}}, \bibnamefont{and}
  \bibinfo{author}{\bibfnamefont{D.}~\bibnamefont{Vanderbilt}},
  \bibinfo{journal}{Reviews of Modern Physics} \textbf{\bibinfo{volume}{84}},
  \bibinfo{pages}{1419} (\bibinfo{year}{2012}).

\bibitem[{\citenamefont{Vaucher et~al.}(2007)\citenamefont{Vaucher, Clark,
  Dorner, and Jaksch}}]{vaucher_fast_2007}
\bibinfo{author}{\bibfnamefont{B.}~\bibnamefont{Vaucher}},
  \bibinfo{author}{\bibfnamefont{S.~R.} \bibnamefont{Clark}},
  \bibinfo{author}{\bibfnamefont{U.}~\bibnamefont{Dorner}}, \bibnamefont{and}
  \bibinfo{author}{\bibfnamefont{D.}~\bibnamefont{Jaksch}},
  \bibinfo{journal}{New Journal of Physics} \textbf{\bibinfo{volume}{9}},
  \bibinfo{pages}{221} (\bibinfo{year}{2007}).

\bibitem[{\citenamefont{Modugno and Pettini}(2012)}]{modugno_maximally_2012}
\bibinfo{author}{\bibfnamefont{M.}~\bibnamefont{Modugno}} \bibnamefont{and}
  \bibinfo{author}{\bibfnamefont{G.}~\bibnamefont{Pettini}},
  \bibinfo{journal}{New Journal of Physics} \textbf{\bibinfo{volume}{14}},
  \bibinfo{pages}{055004} (\bibinfo{year}{2012}).

\bibitem[{\citenamefont{Iba\~nez{-}Azpiroz
  et~al.}(2013{\natexlab{a}})\citenamefont{Iba\~nez{-}Azpiroz, Eiguren,
  Bergara, Pettini, and Modugno}}]{ibanez-azpiroz_self-consistent_2013}
\bibinfo{author}{\bibfnamefont{J.}~\bibnamefont{Iba\~nez{-}Azpiroz}},
  \bibinfo{author}{\bibfnamefont{A.}~\bibnamefont{Eiguren}},
  \bibinfo{author}{\bibfnamefont{A.}~\bibnamefont{Bergara}},
  \bibinfo{author}{\bibfnamefont{G.}~\bibnamefont{Pettini}}, \bibnamefont{and}
  \bibinfo{author}{\bibfnamefont{M.}~\bibnamefont{Modugno}},
  \bibinfo{journal}{Phys. Rev. A} \textbf{\bibinfo{volume}{88}},
  \bibinfo{pages}{033631} (\bibinfo{year}{2013}{\natexlab{a}}).

\bibitem[{\citenamefont{Walters et~al.}(2013)\citenamefont{Walters, Cotugno,
  Johnson, Clark, and Jaksch}}]{walters_textitab_2013}
\bibinfo{author}{\bibfnamefont{R.}~\bibnamefont{Walters}},
  \bibinfo{author}{\bibfnamefont{G.}~\bibnamefont{Cotugno}},
  \bibinfo{author}{\bibfnamefont{T.~H.} \bibnamefont{Johnson}},
  \bibinfo{author}{\bibfnamefont{S.~R.} \bibnamefont{Clark}}, \bibnamefont{and}
  \bibinfo{author}{\bibfnamefont{D.}~\bibnamefont{Jaksch}},
  \bibinfo{journal}{Physical Review A} \textbf{\bibinfo{volume}{87}},
  \bibinfo{pages}{043613} (\bibinfo{year}{2013}).

\bibitem[{\citenamefont{Iba\~nez{-}Azpiroz
  et~al.}(2013{\natexlab{b}})\citenamefont{Iba\~nez{-}Azpiroz, Eiguren,
  Bergara, Pettini, and Modugno}}]{ibanez-azpiroz_tight-binding_2013}
\bibinfo{author}{\bibfnamefont{J.}~\bibnamefont{Iba\~nez{-}Azpiroz}},
  \bibinfo{author}{\bibfnamefont{A.}~\bibnamefont{Eiguren}},
  \bibinfo{author}{\bibfnamefont{A.}~\bibnamefont{Bergara}},
  \bibinfo{author}{\bibfnamefont{G.}~\bibnamefont{Pettini}}, \bibnamefont{and}
  \bibinfo{author}{\bibfnamefont{M.}~\bibnamefont{Modugno}},
  \bibinfo{journal}{Phys. Rev. A} \textbf{\bibinfo{volume}{87}},
  \bibinfo{pages}{011602} (\bibinfo{year}{2013}{\natexlab{b}}).

\bibitem[{\citenamefont{L\"uhmann et~al.}(2014)\citenamefont{L\"uhmann,
  J\"urgensen, Weinberg, Simonet, Soltan-Panahi, and
  Sengstock}}]{luhmann_quantum_2014}
\bibinfo{author}{\bibfnamefont{D.-S.} \bibnamefont{L\"uhmann}},
  \bibinfo{author}{\bibfnamefont{O.}~\bibnamefont{J\"urgensen}},
  \bibinfo{author}{\bibfnamefont{M.}~\bibnamefont{Weinberg}},
  \bibinfo{author}{\bibfnamefont{J.}~\bibnamefont{Simonet}},
  \bibinfo{author}{\bibfnamefont{P.}~\bibnamefont{Soltan-Panahi}},
  \bibnamefont{and}
  \bibinfo{author}{\bibfnamefont{K.}~\bibnamefont{Sengstock}},
  \bibinfo{journal}{Physical Review A} \textbf{\bibinfo{volume}{90}},
  \bibinfo{pages}{013614} (\bibinfo{year}{2014}).

\bibitem[{\citenamefont{Colbert and Miller}(1992)}]{colbert_novel_1992}
\bibinfo{author}{\bibfnamefont{D.~T.} \bibnamefont{Colbert}} \bibnamefont{and}
  \bibinfo{author}{\bibfnamefont{W.~H.} \bibnamefont{Miller}},
  \bibinfo{journal}{The Journal of Chemical Physics}
  \textbf{\bibinfo{volume}{96}}, \bibinfo{pages}{1982} (\bibinfo{year}{1992}).

\bibitem[{\citenamefont{Szalay}(1993)}]{szalay_discrete_1993}
\bibinfo{author}{\bibfnamefont{V.}~\bibnamefont{Szalay}}, \bibinfo{journal}{The
  Journal of Chemical Physics} \textbf{\bibinfo{volume}{99}},
  \bibinfo{pages}{1978} (\bibinfo{year}{1993}).

\bibitem[{\citenamefont{Tiesinga et~al.}(1998)\citenamefont{Tiesinga, Williams,
  and Julienne}}]{tiesinga_photoassociative_1998}
\bibinfo{author}{\bibfnamefont{E.}~\bibnamefont{Tiesinga}},
  \bibinfo{author}{\bibfnamefont{C.~J.} \bibnamefont{Williams}},
  \bibnamefont{and} \bibinfo{author}{\bibfnamefont{P.~S.}
  \bibnamefont{Julienne}}, \bibinfo{journal}{Physical Review A}
  \textbf{\bibinfo{volume}{57}}, \bibinfo{pages}{4257} (\bibinfo{year}{1998}).

\bibitem[{\citenamefont{Littlejohn et~al.}(2002)\citenamefont{Littlejohn,
  Cargo, Jr, Mitchell, and Poirier}}]{littlejohn_general_2002}
\bibinfo{author}{\bibfnamefont{R.~G.} \bibnamefont{Littlejohn}},
  \bibinfo{author}{\bibfnamefont{M.}~\bibnamefont{Cargo}},
  \bibinfo{author}{\bibfnamefont{T.~C.} \bibnamefont{Jr}},
  \bibinfo{author}{\bibfnamefont{K.~A.} \bibnamefont{Mitchell}},
  \bibnamefont{and} \bibinfo{author}{\bibfnamefont{B.}~\bibnamefont{Poirier}},
  \bibinfo{journal}{The Journal of Chemical Physics}
  \textbf{\bibinfo{volume}{116}}, \bibinfo{pages}{8691} (\bibinfo{year}{2002}).

\bibitem[{\citenamefont{Szalay et~al.}(2003)\citenamefont{Szalay, Czak\'o,
  Nagy, Furtenbacher, and Cs\'asz\'ar}}]{szalay_one-dimensional_2003}
\bibinfo{author}{\bibfnamefont{V.}~\bibnamefont{Szalay}},
  \bibinfo{author}{\bibfnamefont{G.}~\bibnamefont{Czak\'o}},
  \bibinfo{author}{\bibfnamefont{A.}~\bibnamefont{Nagy}},
  \bibinfo{author}{\bibfnamefont{T.}~\bibnamefont{Furtenbacher}},
  \bibnamefont{and} \bibinfo{author}{\bibfnamefont{A.~G.}
  \bibnamefont{Cs\'asz\'ar}}, \bibinfo{journal}{The Journal of Chemical
  Physics} \textbf{\bibinfo{volume}{119}}, \bibinfo{pages}{10512}
  (\bibinfo{year}{2003}).

\bibitem[{\citenamefont{Nygaard et~al.}(2004)\citenamefont{Nygaard, Bruun,
  Schneider, Clark, and Feder}}]{nygaard_vortex_2004}
\bibinfo{author}{\bibfnamefont{N.}~\bibnamefont{Nygaard}},
  \bibinfo{author}{\bibfnamefont{G.~M.} \bibnamefont{Bruun}},
  \bibinfo{author}{\bibfnamefont{B.~I.} \bibnamefont{Schneider}},
  \bibinfo{author}{\bibfnamefont{C.~W.} \bibnamefont{Clark}}, \bibnamefont{and}
  \bibinfo{author}{\bibfnamefont{D.~L.} \bibnamefont{Feder}},
  \bibinfo{journal}{Physical Review A} \textbf{\bibinfo{volume}{69}},
  \bibinfo{pages}{053622} (\bibinfo{year}{2004}).

\bibitem[{\citenamefont{Wall et~al.}(2015)\citenamefont{Wall, Hazzard, and
  Rey}}]{wall_effective_2015}
\bibinfo{author}{\bibfnamefont{M.~L.} \bibnamefont{Wall}},
  \bibinfo{author}{\bibfnamefont{K.~R.~A.} \bibnamefont{Hazzard}},
  \bibnamefont{and} \bibinfo{author}{\bibfnamefont{A.~M.} \bibnamefont{Rey}},
  \bibinfo{journal}{Phys. Rev. A} \textbf{\bibinfo{volume}{92}},
  \bibinfo{pages}{013610} (\bibinfo{year}{2015}).

\end{thebibliography}
\end{document}